\documentclass[journal]{IEEEtran}
\usepackage{graphics}
\usepackage{amssymb,amsmath}
\usepackage{graphicx}
\usepackage{subfigure}
\usepackage{float}
\usepackage{multirow}
\usepackage{color}
\usepackage{algorithm}
\usepackage{algpseudocode}

\newtheorem{remark}{\indent Remark}

\title{Robust Learning Control Design for Quantum Unitary Transformations}

\author{Chengzhi Wu, Bo Qi, Chunlin Chen, Daoyi Dong
\thanks{This work was supported by the National Natural Science Foundation of China
(Nos. 61273327, 61374092 and 61432008) and the Australian Research
Council's Discovery Projects funding scheme under Project
DP130101658.}
\thanks{C. Wu is with the Department of Control and Systems Engineering, School of Management and Engineering, Nanjing
University, Nanjing 210093, China (e-mail: stevenczwu@163.com).}
\thanks{B. Qi is with the Key Laboratory of Systems and Control, Academy of Mathematics and Systems Science, Chinese Academy of Sciences, Beijing 100190, China and with the University of Chinese Academy of Sciences, Beijing 100049, China (e-mail: qibo@amss.ac.cn).}
\thanks{C. Chen is with the Department of Control and Systems Engineering, School of Management and Engineering, Nanjing
University, Nanjing 210093, China and with the Research Center for
Novel Technology of Intelligent Equipments, Nanjing University,
Nanjing 210093, China (e-mail: clchen@nju.edu.cn).}
\thanks{D. Dong is with the School of Engineering and Information
Technology, University of New South Wales, Canberra, ACT 2600,
Australia (email: daoyidong@gmail.com).}}

\begin{document}

\maketitle

\begin{abstract}
Robust control design for quantum unitary transformations has been
recognized as a fundamental and challenging task in the
development of quantum information processing due to unavoidable
decoherence or operational errors in the experimental
implementation of quantum operations. In this paper, we extend the
systematic methodology of sampling-based learning control (SLC)
approach with a gradient flow algorithm for the design of robust
quantum unitary transformations. The SLC approach first uses a
``training'' process to find an optimal control strategy robust against certain ranges of
uncertainties. Then a number of randomly selected samples are
tested and the performance is evaluated according to their average
fidelity. The approach is applied to three typical
examples of robust quantum transformation problems including
robust quantum transformations in a three-level quantum
system, in a superconducting
quantum circuit, and in a spin chain
system. Numerical results demonstrate the effectiveness of the
SLC approach and show its potential applications in
various implementation of quantum unitary transformations.
\end{abstract}


\section{Introduction}\label{Sec1}
Quantum information and quantum computation provides the
possibility to run algorithms and protocols superior to those of
its classical counterparts \cite{Bennett 2000Nature}-\cite{Kuang TCYB}.
The methodology of optimal control theory has been
applied in various quantum systems to achieve different
goals, such as attaining a target state, or implementing a desired
quantum gate. As a basic yet influential part in quantum technology,
the generation of unitary transformations is indispensable to the
quantum information processing because the computation carried out
in the quantum logic gates are represented by unitary
transformations. Since the remarkable framework of quantum Turing
machine was proposed by Deutsch \cite{Deutsch 1985PRSLA} in 1985
and the astonishing quantum order-finding algorithm was
announced by Shor \cite{Shor 1994IEEE} in 1994, the research
concerning quantum information processing has blossomed. Numerous
promising candidates for physical implementation of quantum
systems have been proposed in recent decades, such as trapped ions
\cite{Noh 2012NJP}, \cite{Nebendahl 2009PRA}, cavity quantum
electrodynamics (QED) \cite{Zheng 2012PRA}, \cite{Shu 2007PRA},
nuclear magnetic resonance (NMR) \cite{Jones 2011PNMRS},
superconducting qubits based on Josephson junctions \cite{Dong
2015SR}-\cite{Tian 2009IEEE}, and quantum dot in the semiconductor
nanostructures \cite{Kodriano 2014SST}.

Due to the unavoidable existence of imperfection and
uncertainties in the construction of these quantum computation
architectures, it is imperative to design robust implementation
strategies. For example, in NMR, a spin ensemble consisting of
around $10^{23}$ particles are utilized to perform quantum
information processing. The chemical shift of their spectrometers
may not be known exactly \cite{Jones 2011PNMRS}. In the
application of QED, when exciting atoms with lasers to high-lying
Rydberg states or exploiting the long-range dipole-dipole
interaction between Rydberg states, noises are unavoidable in the
microwave coplanar waveguide resonators \cite{Zheng 2012PRA}. In
trapped ions, the bichromatic laser beams may slightly interfere
with each other \cite{Noh 2012NJP}. It is also common that the
temperature may influence the polarization control achieved by
using liquid crystal variable retarders in the semiconductor
quantum dots \cite{Kodriano 2014SST}. Moreover, operations
with multiple superconducting qubits may also confront with the
possible fluctuations in the coupling energy of a Josephson
junction \cite{Dong 2015SR}, \cite{Bialczak 2011PRL}. Hence, it is
both theoretically and practically important to develop
systematic approaches for robust control design for these quantum systems. Lots of
work has been done concerning this problem.
For example,
%Wallraff \emph{et al.} \cite{Wallraff 2005PRL} formulated
%and solved a robust measurement problem of the qubit state by
%coupling the qubit nonresonantly to a transmission line resonator
%and probing the resonator transmission spectrum.
a noise filtering
method has been presented to enhance robustness in quantum
control \cite{Soare 2014NP}. A comprehensive approach of
modulation schemes has been introduced and applied to deal with
the amplitude or phase noise arising from a thermal bath in
two-level systems \cite{Gordon 2007JPB}. Zhang \emph{et al.} \cite{Zhang 2014IEEE} used the
idea of sampling uncertainty parameters to design robust control pulses for electron shuttling.
 A sliding model control scheme has been adopted to deal with
uncertainties in two-level quantum systems \cite{Dong 2009NJP}-\cite{Dong 2012Automatica2}.
In addition, stimulated
Raman adiabatic passage has been extensively studied for its
independence of the pulse shape, which makes it robust against
moderate fluctuations in the experimental parameters \cite{Chen
2009OC}-\cite{Boscain 2012IEEE}.

On the other hand, several methods have been proposed to deal with
the robust transformation problems in quantum systems. In
\cite{Hocker 2014PRA}, the effect of field noise upon target
unitary transformations has been analyzed by investigating the
spectral relationship between the Hessian and the noise in a
general manner.
%Dynamically corrected gates recently have been
%introduced as a tool to achieve decoherence-protected quantum
%gates based on open-loop Hamiltonian engineering \cite{Khodjasteh
%2009PRL}, \cite{Khodjasteh 2009PRA}.
The relationship between
control time and robustness of a quantum system under the
influence of additive white noise has been studied in
\cite{Tibbetts 2012PRA}, and the tradeoff between the fidelity
and its control time has been explored. A  robust optimal control
landscape for the generation of quantum unitary transformations
has been proposed by studying the topology
of the critical regions and Grassmannian submanifolds \cite{Hsieh 2008PRA}.

In classical (non-quantum) engineering, feedback control is usually the preferred choice for robust control, which is also utilized in quantum engineering \cite{Zhang et al 2010}-\cite{Sayrin 2011}. For instance, feedback control with quantum measurements has been used in experiments on single photons
\cite{Gillett 2010PRL} and spin ensembles
\cite{Hammerer 2010RMP}. Nevertheless, considering the small time
scales and the issues resulting from the measurement backaction in
physical quantum systems, open-loop control is a more practical
choice with the current level of quantum technology. For example,
Khaneja \emph{et al.} \cite{Khanejia 2005JMR} designed optimal
control pulse sequences for NMR by applying gradient algorithms.
Kosut \emph{et al.} \cite{Koust 2013PRA} proposed a sequential
convex programming method for designing robust quantum
manipulations. However, it is still a challenging
task to provide a practical approach for the implementation of
general quantum unitary transformations.

In this paper, we employ the sampling-based learning control (SLC)
method for designing robust quantum unitary transformations. The
SLC approach was originally presented for control design of
inhomogeneous quantum ensembles \cite{Chen 2014PRA} and robust
control of quantum states \cite{Dong 2015IEEE}, \cite{Dong
2013IEEE}. The approach has also been applied for the robust
manipulation of superconducting qubits and a set of quantum gates
in the presence of fluctuations \cite{Dong 2015SR}, \cite{Dong
2016arXiv} and quantum ensemble classification \cite{quantum
classification}. The SLC method includes two steps of ``training''
and ``testing''. In the training step, several samples are
selected according to the distribution of Hamiltonian
uncertainties to construct an augmented system, then a gradient
flow based learning and optimization algorithm is applied on the
augmented system to find a robust control law for a desired
quantum unitary transformation. In the process of testing step, we
apply the control law obtained in the ``training'' step to a
number of samples whose parameters are selected according to
uniform distribution or truncated Gaussian distribution, and
evaluate the performance according to their average fidelity.

We demonstrate the application of the SLC
method to three typical examples of quantum unitary transformations. In these examples, we assume that there exist some uncertainties and aim at designing robust control fields that can achieve high-fidelity quantum unitary transformations. For simplicity, we consider time-invariant uncertainties in these numerical examples. These results are straightforwardly applicable to time-varying uncertainties. The first example is a quantum
unitary transformation problem in a three-level quantum
system, which is a widely used model in natural and artificial
ions and atoms. The second example is concerning the
superconducting quantum circuits, which have been proved to be one
of the promising alternatives for quantum information processing
\cite{Clarke 2008Nature}-\cite{Hofheinz 2009Nature}. In particular,
different robust quantum transformations including SWAP, CPhase and
CHadamard are implemented using SLC. In the third example, we
investigate the application of SLC in a spin chain system, which
has been widely studied and applied due to its long decoherence
and relaxation time \cite{Heule 2010PRA}-\cite{Konstantinidis
2015JP}. Numerical results show that the SLC method is
effective for robust control design of quantum unitary
transformations.

The rest of the paper is organized as follows. Section \ref{Sec2}
formulates the control problem of quantum unitary transformations.
Section \ref{Sec3} introduce the approach of sampling-based
learning control and a gradient flow based learning and
optimization algorithm.
The results of robust control design for quantum unitary
transformations in a three-level quantum system is presented in
Section \ref{Sec4}. Section \ref{Sec5} demonstrates the
application of the SLC approach to achieve three robust quantum
transformations (i.e., SWAP, CPhase and CHadamard operations) in
quantum superconducting circuits. In Section \ref{Sec6}, the SLC
approach is utilized to learn a robust control law for a spin
chain of Heisenberg XXX model. Conclusions are presented in
Section \ref{Sec7}.

\section{Problem formulation}\label{Sec2}
For a quantum system, if it can be approximated as a closed system,
its state may be described by a complex vector $(\alpha_1,\alpha_2,\cdots)^{T}$,
where $\alpha_j\ (j=1,2,\cdots)$ are complex numbers satisfying
$\sum_j |\alpha_j|^2=1$. In this paper, we consider finite-dimensional
systems and assume the dimension as $D$. An operation on
a quantum system can be described by a unitary transformation
$U$ which turns the system from an initial state $(\alpha_1,\alpha_2,\cdots,\alpha_D)^{T}$
to another state $(\beta_1,\beta_2,\cdots,\beta_D)^{T}$, i.e.,
\begin{equation}
(\beta_1,\beta_2,\cdots,\beta_D)^{T}= U (\alpha_1,\alpha_2,\cdots,\alpha_D)^{T},
\end{equation}
where the unitary transformation $U$ is a $D\times D$ complex
matrix satisfying $U^\dag U=I$, and $U^\dag$ is the conjugate
transpose of $U$.

Denoted in the matrix form, the controlled
evolution of a unitary transformation $U$ on a
quantum system satisfies
\begin{equation}\label{systemequation}
\left\{ \begin{split}%{c}
  \frac{d}{dt}{U}(t) &=-iH(t)U(t)\\
             &=-i\{H_0+\sum_{m=1}^{M}u_m(t)H_m\}U(t),\\
  t \in [&0, T],\\
  \ U(0)&=I,\\
\end{split}
\right.
\end{equation}
where $H_0$ is the free Hamiltonian of the system, $H_m\ (m=1,2,\cdots M)$
are the control Hamiltonians, and $u_m(t)$ are the corresponding control
pulse sequences. Here we use atomic units (a.u.) by setting the
reduced Planck constant $\hbar=1$. At $t=0$, the initial
unitary transformation $U(t)$ is $U(0)=I$. Then we have the
propagator at the final time $T$
\begin{equation}
U(T)=\mathbb{T} \text{exp}(-i\int_0^T H(t) dt),
\end{equation}
where $\mathbb{T}$ is the time-ordering operator. In practical
applications, the transfer time $T$ is usually discretized
into $N$ equal intervals, and the control amplitudes keep constant
during each interval. Let $U_j$ denote the unitary transformation during
the $j$th interval. Thus this quantum transformation can be divided
into $N$ smaller pieces $U_1, U_2, \dots , U_N$, which satisfy
$U_N \cdots U_2 U_1=U(T)$. Hence, the unitary transformation of
the system at $j$th interval is
\begin{equation}
U_{j}=\text{exp}(-i\int_{(j-1)\Delta t}^{j\Delta t} H(t) dt),
\end{equation}
where $\Delta t=T/N$ is the time interval.

For realistic quantum systems, there may be uncertainties in the control
fields or the coupling strength. In this paper, we consider the
uncertainties that can be formulated in the system Hamiltonian $H(t)$ as
\begin{equation}
H(t)=f_0(\varepsilon_{0})H_{0}+\sum_{m=1}^{M}f_m(\varepsilon_{m})u_{m}(t)H_{m},
\end{equation}
where $f_0(\varepsilon_{0})$ and $f_m(\varepsilon_{m})$
characterize these uncertainties. We assume that
$\varepsilon_{0} \in [1-E_{0}, 1+E_{0}]$ and $\varepsilon_{m} \in
[1-E_{m}, 1+E_{m}]$ are time independent. The constants $E_{0}\in
[0,1]$ and $E_{m} \in [0,1]$ represent the normalized bounds of
the uncertainty parameters. The aim is to design a control
strategy $\{u_{m}(t), m=1,2,\ldots , M\}$ to achieve robust
manipulation, i.e., to construct the unitary $U(t)$ starting from
the initial transformation $U(0)=I$ to the target transformation $U_F$ with a high
level of fidelity in the presence of uncertainties. In this paper,
the fidelity is defined as
\begin{equation}
\begin{split}
F(U_F,U(T)) &= \frac{1}{D} |\langle U_F|U(T)\rangle|  \\
            &= \frac{1}{D} |\text{tr}\{U_F^\dagger U(T)\}| .  \\
\end{split}
\end{equation}
This fidelity is often used to measure the difference between
two unitary transformations \cite{Palao 2002PRL}, \cite{Thomas 2011PRA}.

The control performance can be described by a \emph{performance
function} $J(u)$ for each control strategy $u=\{u_{m}(j),
m=1,2,\ldots , M, j=1,2,\ldots , N\}$ and the control problem can
be formulated as a maximization problem as follows:
\begin{equation}
\begin{split}
\max\limits_u \ \ \  & J(u):= \max\limits_u \ \mathbb{A} \ [\ \frac{1}{D} |\text{tr}\{U_F^\dagger U(T)\}|\ ] \\
\text{s.t.} \ \ \ & \frac{d}{dt}{U}(t)=-iH(t)U(t), \\
& H(t)=f_0(\varepsilon_{0})H_{0}+\sum_{m=1}^{M}f_m(\varepsilon_{m})u_{m}(t)H_{m}, \\
& t\in [0, T], \\
& U(0)=I, \\
& \varepsilon_{0} \in [1-E_{0}, 1+E_{0}], \\
& \varepsilon_{m} \in[1-E_{m}, 1+E_{m}],
\end{split}
\end{equation}
where $J(u)$ is dependent on the control fields $u$ through the
Schr\"{o}dinger equation and $\mathbb{A}[\cdot]$ denotes the
average expectation fidelity with uncertainty parameters
$\mathcal{E}=\{\varepsilon_{0},\varepsilon_{m}, m=1,2,\cdots,M\}$.

\begin{figure}
\centering
\includegraphics[width=0.5\textwidth]{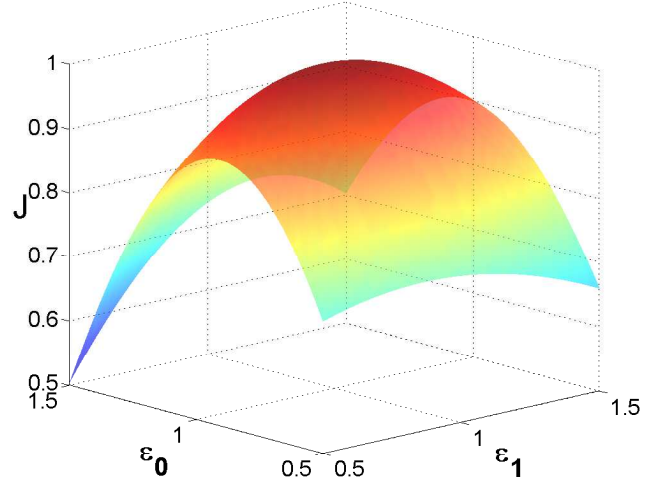}
\caption{Two uncertainty parameters $\varepsilon_0$ and $\varepsilon_1$ are considered in the example of Hadamard operation for the two-level quantum system. With a non-robust control strategy, the fidelity of the system is miserable when its uncertainty parameters fluctuate, which shows that the control strategy is of poor robustness.}\label{nonrobust}
\end{figure}

Now, we take the Hadamard operation on a two-level quantum system
in \cite{Wu 2015SMC} as an example to demonstrate the demand of
robust control design. The Hadamard operation is a special
unitary transformation denoted as
\begin{equation}
\text{Hadamard}=\frac{1}{\sqrt{2}}\begin{pmatrix}
  1 & 1  \\
  1 & -1  \\
\end{pmatrix}.
\end{equation}
Here, its system Hamiltonian is
denoted by $H(t)=f_0(\varepsilon_{0})u_{0}\sigma_{z}+f_1(\varepsilon_{1})u_{x}(t)\sigma_{x}$,
where $\varepsilon_0$ and $\varepsilon_1$ are its uncertainty
parameters on the free Hamiltonian and the control Hamiltonian, respectively. We first find an optimal control law for the nominal system without
parameter fluctuations. Fig. \ref{nonrobust} shows the fidelities
that this system may achieve as its parameters are fluctuating
using the optimal control law. With this non-robust control strategy $u^*=(u_0^*, u_x^*)$, the
fidelity of the system declines to 0.8 when parameter fluctuations
reach 20\%. Moreover, when parameter fluctuations grow to 40\%, the
fidelity drops to 0.6. Hence, it is of vital importance to develop
a general approach to achieve robust performance.

\section{Robust learning control design method}\label{Sec3}

To design a robust control law for the implementation of quantum unitary transformations, we develop a systematic numerical algorithm within the framework of SLC.
The SLC method includes two steps of ``training'' and ``testing'' \cite{Chen 2014PRA}.
In the training step, we select $X$ samples to train the control fields. These samples are selected according to the possible distribution of Hamiltonian uncertainties (e.g., uniform distribution). Let $\overrightarrow{\varepsilon_{i}}=(\varepsilon_{1i}, \varepsilon_{2i}, \ldots, \varepsilon_{Mi})$, and an augmented system can be constructed using these samples as
\begin{equation}
\left(
\begin{array}{c}
  \frac{d}{dt}{U}_{\varepsilon_{01},\overrightarrow{\varepsilon_{1}}}(t) \\
  \frac{d}{dt}{U}_{\varepsilon_{02},\overrightarrow{\varepsilon_{2}}}(t) \\
  \vdots \\
%  \dot{U}_{\varepsilon_{0X},\stackrel{\rightharpoonup}{\varepsilon_{X}}}(t) \\
  \frac{d}{dt}{U}_{\varepsilon_{0\!X},\overrightarrow{\varepsilon_{\!X}}}(t) \\
\end{array}
\right)= -i
\left(
\begin{array}{c}
  {H}_{\varepsilon_{01},\overrightarrow{\varepsilon_{1}}}(t){U}_{\varepsilon_{01},\overrightarrow{\varepsilon_{1}}}(t) \\
  {H}_{\varepsilon_{02},\overrightarrow{\varepsilon_{2}}}(t){U}_{\varepsilon_{02},\overrightarrow{\varepsilon_{2}}}(t) \\
  \vdots \\
  {H}_{\varepsilon_{0\!X},\overrightarrow{\varepsilon_{\!X}}}(t){U}_{\varepsilon_{0\!X},\overrightarrow{\varepsilon_{\!X}}}(t) \\
\end{array}
\right),
\end{equation}
where $H_{\varepsilon_{0i},\overrightarrow{\varepsilon_{i}}}(t)=f_0(\varepsilon_{0i})H_{0}+\sum_{m=1}^{M}f_m(\varepsilon_{mi})u_{m}(t)H_{m}$ with $i=1,2,\ldots,X$. The average performance index of the augmented system is defined as
\begin{equation}
F_X(u)=\frac{1}{X} \sum_{i=1}^{X} \frac{1}{D} |\langle U_F|U_{\varepsilon_{0i},\overrightarrow{\varepsilon_{i}}}(T)\rangle|.
\end{equation}
The task of the training step is to find an optimal control field $u^*=(u_1,u_2,\ldots,u_M)^*$ to maximize the performance index $F_X(u)$.

In order to obtain good performance, it is necessary to choose representative samples for the uncertainties. If the distributions of parameters are uniform, the intervals $[-E_0,E_0]$ and $[-E_m,E_m]$ can be divided into $N_{0}+1$ and $N_{m}+1$ subintervals, then we select $N_{0}$ samples for $\varepsilon_{0}$ and $N_{m}$ samples for $\varepsilon_{m}$, i.e., $X=N_{0}\prod_{m=1}^{M}N_{m}$ samples in total. Usually, larger $N_{0}$ and $N_{m}$ could lead to better performance. However, they require longer computational time. In this paper, we select 5 samples for each uncertainty parameter, which can achieve good performance \cite{Dong 2015SR}. $\varepsilon_{0i}$ and $\overrightarrow{\varepsilon_{i}}$ can be chosen from the combination of $(\varepsilon_{0i},\varepsilon_{1i},\ldots,\varepsilon_{Mi})$ as follows:
\begin{equation}
\left\{ \begin{split}
& \varepsilon_{0i}\in\{1-E_0+\frac{(2i_{0}-1)E_0}{N_{0}},i_{0}=1,2,\ldots,N_{0}\} \\
& \varepsilon_{mi}\!\in\!\{1\!-\!E_m\!+\!\frac{(2i_{m}-1)E_m}{N_{m}},i_{m}\!=\!1,2,\ldots,N_{m}\}. \\
\end{split}
\right.
\end{equation}

In the testing step, we apply the control fields $u^*$ obtained in the training step to a large number of additional samples, which are randomly selected according to the parameter uncertainties. Each sample will be evaluated to get its performance. If the average fidelity of all the tested samples is satisfactory, the learned control strategy is acceptable and the quantum unitary transformation we achieved is robust. In this paper, we use 1000 samples to test the learned control strategy in the testing step.

Within the SLC framework, a key task is to develop an appropriate algorithm to solve the optimization problem in the training step. The gradient flow method has shown to be one of the most efficient methods to solve optimal control problems. In this paper, we consider the problem of realizing a desired quantum unitary transformation with a high level of fidelity within a given time $T$. The gradient flow algorithm is described as follows.

Let $U_F$ denote the desired target quantum transformation,
the performance index can be defined by the Hilbert-Schmidt distance between the target unitary transformation $U_F$ and the controlled unitary transformation $U(T)$ as
\begin{equation}
\|U_F-U(T)\|^2=\|U_F\|^2-2Re\langle U_F|U(T)\rangle+\|U(T)\|^2.
\end{equation}
For practical applications, considering the existence of an arbitrary global phase factor $e^{i\varphi}$,
the problem is changed to minimize
\begin{equation}
\|U_F-e^{i\varphi}U(T)\|^2=\|U_F\|^2-2Re\langle U_F|e^{i\varphi}U(T)\rangle+\|e^{i\varphi}U(T)\|^2,
\end{equation}
which is equivalent to maximize $Re\langle U_F|e^{i\varphi}U(T)\rangle$.
It can be verified that this problem is equivalent to maximize
\begin{equation}\label{performance function}
\Phi=|{\langle U_F|e^{i\varphi}U(T)\rangle}|^2.
\end{equation}

With operators $A_j$ and $B_j$ defined as $A_j=U_j\cdots U_1$ and
$B_j=U_{j+1}^\dagger \cdots U_N^\dagger U_F=A_jU(T)^\dagger U_F$,
we can derive the performance function as
{\setlength\arraycolsep{1.3pt}
\begin{eqnarray}
\Phi&=&|{\langle U_F|e^{i\varphi}U(T)\rangle}|^2
\nonumber\\
&=&\langle U_F|e^{i\varphi}U_N\cdots U_1\rangle \langle e^{i\varphi}U_N\cdots U_1|U_F\rangle
\nonumber\\
&=&\langle U_{j+1}^\dagger \cdots U_N^\dagger U_F|U_j\cdots U_1\rangle
\langle U_j\cdots U_1|U_{j+1}^\dagger \cdots U_N^\dagger U_F\rangle
\nonumber\\
&=&\langle B_j|A_j\rangle \langle A_j|B_j\rangle.
\end{eqnarray}
%\begin{equation}       %another format
%\begin{split}
%\Phi &= |{\langle U_F|e^{i\varphi}U(T)\rangle}|^2  \\
%     &= \langle U_F|e^{i\varphi}U_N\cdots U_1\rangle \langle e^{i\varphi}U_N\cdots U_1|U_F\rangle  \\
%     &= \langle U_{j+1}^\dagger \cdots U_N^\dagger U_F|U_j\cdots U_1\rangle
%        \langle U_j\cdots U_1|U_{j+1}^\dagger \cdots U_N^\dagger U_F\rangle  \\
%     &= \langle B_j|A_j\rangle \langle A_j|B_j\rangle.  \\
%\end{split}
%\end{equation}
Let us see how the performance $\Phi$ changes when we perturb the control amplitude from $u_m(j)$ to $u_m(j)+\delta u_m(j)$ at step $j$. According to Eq. (\ref{systemequation}), we have
\begin{equation}\label{gradeint of U_j1}
\begin{split}
\frac{d}{dt}(U_j(t)+\delta U_j(t)) &= -i\{H_0+\sum_{m=1}^{M}u_m(t)H_m  \\
                           & \quad +\delta u_m(j)H_m\}(U_j(t)+\delta U_j(t)).
\end{split}
\end{equation}
Compared with Eq. (\ref{systemequation}), Eq. (\ref{gradeint of U_j1}) becomes
\begin{equation}\label{gradeint of U_j2}
\frac{d}{dt}{\delta U_j(t)} = -i\{H_0+\sum_{m=1}^{M}u_m(t)H_m\}\delta U_j(t)-i\delta u_m(j)H_m U_j(t).
\end{equation}
By ignoring higher order terms, we have
\begin{equation}\label{gradeint of U_j3}
\delta U_j(\Delta t) = \int_0^{\Delta t} -i\delta u_m(j)U_j(\Delta t-\tau)H_mU_j(\tau) d\tau.
\end{equation}
Employing the standard formula
\begin{equation}
e^{-SA}Be^{SA} = B-S[A,B]+\frac{s^2}{2!}[A,[A,B]]+\cdots,
\end{equation}
Eq. (\ref{gradeint of U_j3}) becomes
\begin{small}
\begin{equation}
\begin{split}
\delta U_j(\Delta t) &= -i\delta u_m(j)(\int_0^{\Delta t}\!U_j(\Delta t-\tau)H_mU_j(\tau-\Delta t) d\tau) U_j(\Delta t)  \\
                     &= -i\delta u_m(j)(H_m\Delta t + \frac{(\Delta t)^2}{2}i[H(j),H_m] \!+\!\cdots) U_j(\Delta t).
\end{split}
\end{equation}
\end{small}
Considering the first order approximation, we can obtain the change in $U_j$ as
\begin{equation}
\delta U_j(\Delta t) = -i\Delta t\delta u_m(j)H_m U_j(\Delta t).
\end{equation}
Therefore, the corresponding gradient $\delta\Phi/\delta u_m(j)$ to first order in $\Delta t$ is given by
{\setlength\arraycolsep{1.3pt}
\begin{small}
\begin{eqnarray}
\frac{\delta\Phi}{\delta u_m(j)}&=&-\langle B_j|i\Delta t H_m A_j\rangle \langle A_j|B_j\rangle
-\langle B_j|A_j\rangle \langle i\Delta t H_m A_j|B_j\rangle
\nonumber\\
&=&-2Re\{\langle B_j|i\Delta t H_m A_j\rangle \langle A_j|B_j\rangle\},
\end{eqnarray}
\end{small}
where $u_m(j)$ is the amplitude of $u_m$ at the $j$th step.

During the iteration, the control fields can be updated according to the law:
\begin{equation}
u_m^{k+1}(j)=u_m^k(j)+\alpha_s \frac{\delta\Phi}{\delta u_m(j)},
\end{equation}
where $\alpha_s$ is the step size.

\begin{remark}
It is clear that in the iterative updating process, we try to maximize the performance function $\Phi$ by a forward Euler method. Although as an explicit method, the forward Euler method may be not able to ensure numerical stability and there is a limitation on the step size $\alpha_s$, it is very easy to implement and has low computation cost compared with the implicit backward Euler method. A large number of numerical results of quantum optimization problems show that it is easy to choose a suitable step size to achieve the numerical stability for the forward Euler method.
\end{remark}

\begin{algorithm}
\caption{\ Gradient flow method with sampling-based learning control } \label{Gradient Flow}
\begin{algorithmic}[1]
\State Set the index of iterations $k=0$
\State Choose a set of arbitrary controls $u^{k=0}=\{u_{m}^{0}(j), \ m=1,2,\ldots,M, \ j=1,2,\ldots,N\}$
\Repeat {\ (for each iterative process)}
\Repeat {\ (for each training samples $i=1,2,\ldots,X$)}
\State Compute the propagator $U_{i}^{k}(j)$ with the control strategy $u^{k}$
\State $U_{i}(T,u^{k})=U^{k}_{i}(N) \cdots U^{k}_{i}(1)U_{0}$
\Until {\ $i=X$}
\State Compute the fidelity $F^k$ of these $X$ samples
\State $F^k=\frac{1}{X} \sum_{i=1}^{X} \frac{1}{D} |\langle U_F|U_{i}(T,u^{k})\rangle|$
\Repeat {\ (for each control $u_{m}(m=1,2,\ldots,M)$ of the control field $u$)}
\State \begin{small}$\delta_m^{k}(j)=-\frac{2}{X} \sum^{X}_{i=1}Re\{\langle B_j^i|i\Delta t H_m A_j^i\rangle \langle A_j^i|B_j^i\rangle\}$\end{small}
\State $u_{m}^{k+1}(j)=u_{m}^{k}(j)+\alpha_s\delta_{m}^{k}(j)$
\Until {\ $m=M$}
\State $k=k+1$
\Until {\ the learning process ends}
\State The optimal control strategy
$u^{*}=\{u_{m}^*\}=\{u_{m}^{k}\}, \ m=1,2,\ldots,M$
\end{algorithmic}
\end{algorithm}

\emph{Algorithm 1} gives the algorithm of gradient flow with the SLC approach in the training step.
In \emph{Algorithm 1}, the updating rule (line 10-13) is formalized as a batch gradient descent method to maximize the averaged performance index of the augmented system constructed with $X$ samples. The convergence can be expected as long as the step size $\alpha_s$ is well set, i.e., $\sum_{k=1}^\infty \alpha_{s,k}^2 < \infty$, and $\sum_{k=1}^\infty \alpha_{s,k} = \infty$.
Apart from the traditional gradient ascent/descent algorithm, there are lots of other algorithms in the optimal control theory, such as the conjugate gradient method, and the biconjugate gradient method. There are also some second-order algorithms, like the Newton method or the quasi-Newton method, by using the Hessian. In this paper, the gradient ascent pulse engineering (GRAPE) method \cite{Khanejia 2005JMR} is used due to its simplicity and effectiveness.

\section{Robust quantum transformation in three-level quantum systems}\label{Sec4}
%In this section, the proposed method is applied to achieve robust quantum unitary transformation in a V-type three-level quantum system, which is a typical prototype to lots of quantum systems involving the evolution of atoms.
%It is an essential task because robustness lies at the central problem of turning theoretical quantum models into practical applications.
%For simplicity, we use atomic units (a.u.) in this section, i.e., setting $\hbar=1$.

%In addition, we assume that there are no constraints on the external controls.

    \subsection{The system}
Considering a V-type quantum system, assume that the initial state is $|\psi_1\rangle=(\alpha_1,\beta_1,\gamma_1)^T$ and the target state is $|\psi_2\rangle=(\alpha_2,\beta_2,\gamma_2)^T$, and the target quantum transformation is $U_F$ that satisfies $|\psi_2\rangle=U_F|\psi_1\rangle$. The evolution equation of the quantum transformation is given as
\begin{equation}
\frac{d}{dt}{U(t)}=-i[u_0H_{0}+u_{1}(t)H_{1}+u_{2}(t)H_{2}+u_{3}(t)H_{3}]U(t).
\end{equation}
In this model, we use the Gell-Mann matrices to represent the free and the control Hamiltonians.
Let $H_0=\sigma_3$, $H_1=\sigma_1$, $H_2=\sigma_4$, and $H_3=\sigma_6$, where
\begin{equation}
\sigma_1=\begin{pmatrix}
  0 & \ 1 \ & 0  \\
  1 & 0 & 0  \\
  0 & 0 & 0  \\
\end{pmatrix},
\sigma_3=\begin{pmatrix}
  1 & 0 & 0  \\
  0 & -1 & 0  \\
  0 & 0 & 0  \\
\end{pmatrix}, \nonumber
\end{equation}
\begin{equation}
\sigma_4=\begin{pmatrix}
  0 & \ 0 \ & 1  \\
  0 & 0 & 0  \\
  1 & 0 & 0  \\
\end{pmatrix},
\sigma_6=\begin{pmatrix}
  0 & \ 0 \ & 0  \\
  0 & 0 & 1  \\
  0 & 1 & 0  \\
\end{pmatrix},
\end{equation}
are the Gell-Mann matrices.
Suppose that uncertainties exist in both of the free Hamiltonian and control Hamiltonians, i.e.,
\begin{equation}
H(t)=f_0(\varepsilon_{0})u_0H_{0}+\sum_{m=1}^{3}f_m(\varepsilon_{m})u_{m}(t)H_{m}.
\end{equation}

For simplicity, we assume that $f_0(\varepsilon_{0})=\varepsilon_{0}$ and $f_m(\varepsilon_{m})=f(\varepsilon_1)=\varepsilon_1$ for all $m=1, 2, 3$ and $E_0=E_1=E$, i.e., $\varepsilon_0,\varepsilon_1 \in[1-E, 1+E]$, where $E\in[0, 1]$ is a given constant. $\varepsilon_{0}$ and $\varepsilon_{1}$ are assumed to have uniform distributions in their ranges of fluctuations, respectively. In the training step of SLC, an augmented system is constructed by selecting $N_{0}=5$ for $\varepsilon_{0}$ and $N_{1}=5$ for $\varepsilon_{1}$. The samples are selected as:

\begin{equation}
\left\{ \begin{split}
& \varepsilon_{0}=1-E+\frac{(2\text{fix}((n-1)/5)+1)E}{5},\\
& \varepsilon_{1}=1-E+\frac{(2\text{mod}((n-1),5)+1)E}{5}, \\
\end{split}
\right. \ \ \
\end{equation}
where $n=1,2,\ldots,25$, fix$(X)$ rounds $X$ to the nearest integer towards zero, and mod$(X,Y)$ returns the remainder of the division of $X$ by  $Y$.
In the testing step of SLC, we use 1000 samples for testing.

\begin{figure}
\centering
\includegraphics[width=0.5\textwidth]{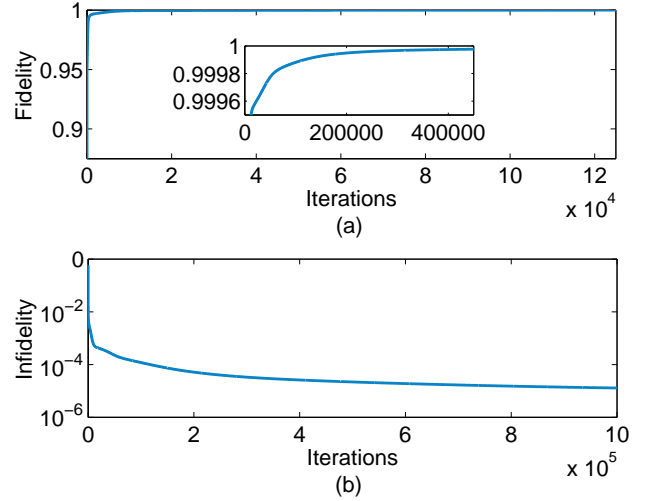}
\caption{The learning performance (a) the fidelity; (b) the infidelity versus the number of iterations for the quantum transformation $U_{3\times 3}$ with uncertainty parameters of $\varepsilon_0$ and $\varepsilon_1$.}\label{levelthree1}
\end{figure}

\begin{figure}
\centering
\includegraphics[width=0.5\textwidth]{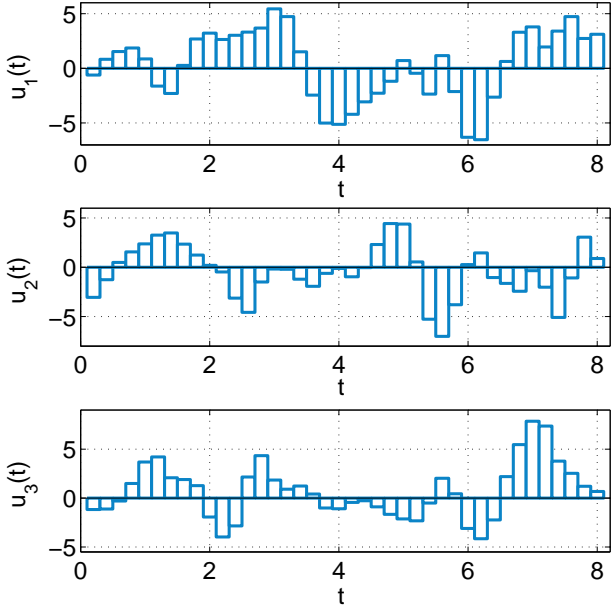}
\caption{The learned control strategy for the quantum transformation $U_{3\times 3}$.}\label{levelthree2}
\end{figure}

\begin{figure}
\centering
\includegraphics[width=0.5\textwidth]{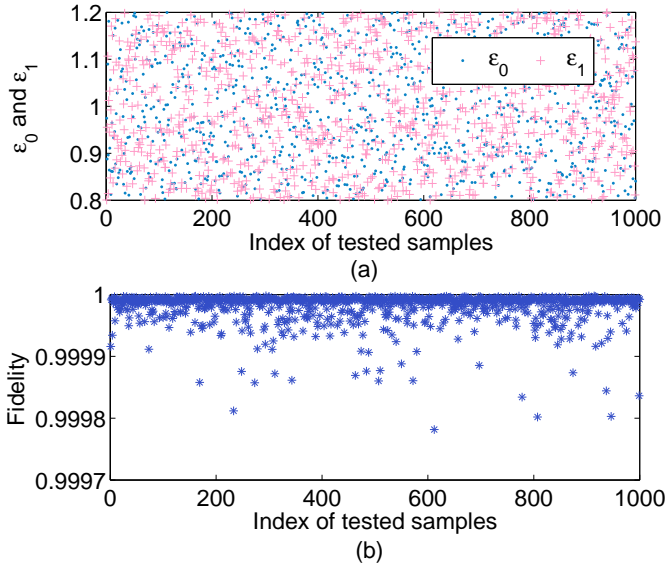}
\caption{The testing performance of the quantum transformation $U_{3\times 3}$: (a) 1000 randomly selected testing samples; (b) the testing performances regarding fidelity.}\label{levelthree3}
\end{figure}

    \subsection{Numerical results}
The quantum unitary transformation on a three-level quantum system can be denoted by a $3\times 3$ unitary matrix $U_{3\times 3}$. In this section, we randomly select one as the target unitary transformation, e.g.,
\begin{equation}
U_{3\times 3}=\begin{pmatrix}
  \ -\frac{1}{\sqrt{3}} & \ \frac{1}{\sqrt{2}} \ & \frac{1}{\sqrt{6}}  \\
  \ -\frac{1}{\sqrt{3}} & 0 & -\frac{2}{\sqrt{6}}  \\
  \ -\frac{1}{\sqrt{3}} & -\frac{1}{\sqrt{2}} & \frac{1}{\sqrt{6}}  \\
\end{pmatrix}.
\end{equation}

Now we use the proposed method to find a robust control sequence to achieve the quantum transformation $U_{3\times 3}$. The infidelity is adopted as the performance index ($\text{Infidelity}=1-\text{Fidelity}$). We assume $T=8$, $u_{0}=1$, $u_{m}\in[-5, 5]$ and approximate each control field using piece-wise pulses that may be easy to implement in some practical quantum systems.
For example, the manipulation time $T$ can be divided into 40 intervals where a constant pulse is applied during each interval. Here we use 40 piece-wise subpulses to consider the tradeoff between the computational cost and the performance. Usually, more subpulses could achieve better performance while more computational cost is required. The default initial control field is $u_m=\sin t$. The boundary of the fluctuations is set as $E=0.2$.
The iteration step size is set as $\alpha_s=0.1$.

The training performance is illustrated in Fig. \ref{levelthree1}, which shows that the average fidelity of the augmented system converges to 0.9999 after 1,000,000 iterations. The learned control strategy is shown as in Fig. \ref{levelthree2}. Then in the testing step, the learned fields are applied to 1000 randomly selected samples whose parameters are chosen according to the uniform distribution. As shown in Fig. \ref{levelthree3}, the average fidelity reaches 0.99998 and shows that the realization of the
quantum transformation under the learned control is of great robustness.

\section{Robust unitary transformation in superconducting circuits}\label{Sec5}
In this section, the proposed approach is applied to some physical systems of superconducting circuits.
The presented method is very flexible in the selection of the initial control $u_m(t)$ and the operation time $T$, as well as the target unitary transformation. It is also robust against fluctuations in different parameters.

\subsection{The physical setup}
Considering the physical realization of quantum computers, solid-state devices may be promising candidates. Among them, superconducting quantum circuits have been widely studied. Based on Josephson junctions, these superconducting circuits can behave quantum mechanically like artificial atoms, offering a promising way for quantum information processing.
Furthermore, superconducting quantum circuits provide efficient solutions for quantum computer architectures when extended to a large number of qubits due to their exceptional ability of scalability, tunability and design flexibility.

Following the early superconducting qubits scheme proposed by Shnirman {\em et al.} \cite{Shnirman 1997PRL} , a series of results of superconducting qubits have been conducted to explore the properties. In superconducting quantum circuits, the Josephson qubit can be achieved in a Cooper-pair box, which is a small superconducting island weakly coupled to a bulk superconductor through a Josephson-Junction and driven by a voltage source through a gate capacitance. In a superconducting qubit, two significant quantities are the Josephson coupling energy $E_J$ and the charging energy $E_C$, whose ratio determines that the behaviour of qubit is dominated by phase or charge \cite{You 2005PT}, \cite{Hofheinz 2009Nature} (see Fig. \ref{SQUIDpic}(a)). When $E_C \gg E_J$, a superconducting charge qubit is constructed, whose Hamiltonian can be described as
\begin{equation}
H=E_C(n-n_g)^2-E_Jcos\phi,
\end{equation}
where the phase drop $\phi$ across the Josephson-Junction is conjugate to the number $n$ of extra Cooper pairs in the box, $n_g=C_gV_g/2e$ is controlled by the external gate voltage $V_g$, $C_g$ is the gate capacitance and $2e$ is the charge of each Cooper pair. In most experiments, in order to get a better control over the qubit, physicists often use a dc superconducting interference device (SQUID) loop instead, which is constructed by two Josephson-Junctions, as shown in Fig. \ref{SQUIDpic}(b). The Hamiltonian of the system can be approximated as
\begin{equation}
H=f(V_g)\sigma_z-g(\Phi)\sigma_x,
\end{equation}
where $f(V_g)$ is relevant to the charging energy $E_C$ and can be adjusted by $V_g$, $g(\Phi)$ is relevant to the coupling energy $E_J$ and can be controlled by the magnetic flux $\Phi$ applied through the SQUID loop.
The Pauli matrices $\sigma=(\sigma_x,\sigma_y,\sigma_z)$ are
\begin{equation}
\sigma_x=\begin{pmatrix}
  0 & \ 1  \\
  1 & \ 0  \\
\end{pmatrix},
\sigma_y=\begin{pmatrix}
  0 & -i  \\
  i & 0  \\
\end{pmatrix},
\sigma_z=\begin{pmatrix}
  1 & 0  \\
  0 & -1  \\
\end{pmatrix}.
\end{equation}

\begin{figure}
\centering
\includegraphics[width=0.45\textwidth]{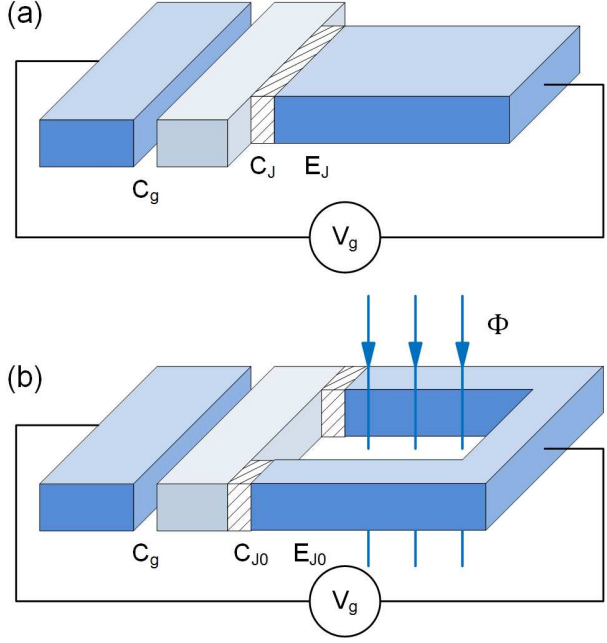}
\caption{Superconducting charge qubit: (a) the Cooper-pair box is connected to another block of superconducting metal through a Josephson junction (the white shaded parts) and driven by a voltage source $V_g$ through a gate capacitance $C_g$. The capacitance and the coupling energy of the Josephson junction are $C_J$ and $E_J$, respectively. (b) the Josephson junction is replaced by a double-junction SQUID loop, and the capacitance and the coupling energy of each Josephson junction are $C_{J0}$ and $E_{J0}$, respectively. In addition, a magnetic flux $\Phi$ is applied through the SQUID loop to control the effective Josephson coupling energy.}\label{SQUIDpic}
\end{figure}

Here, we consider an example of two coupled superconducting qubits \cite{Bialczak 2011PRL}. Each qubit is a nonlinear resonator built from an Al/AlO$_x$/Al Josephson junction, and the two qubits are coupled via a modular four-terminal device. This four-terminal device is constructed using two nontunable inductors, a fixed mutual inductance and a tunable inductance. The equivalent Hamiltonian can be described as \cite{Dong 2015SR}
\begin{equation}
\begin{split}
H &= \frac{\hbar\omega_{1}(t)}{2}\sigma^{(1)}_{z}+\frac{\hbar\omega_{2}(t)}{2}\sigma^{(2)}_{z}
     +\frac{\hbar\omega_{3}(t)}{2}\sigma^{(1)}_{x}+\frac{\hbar \omega_{4}(t)}{2}\sigma^{(2)}_{x}  \\
  & \quad +\frac{\hbar\Omega_{c}(t)}{2}(\sigma^{(1)}_{x}\sigma^{(2)}_{x}
    +\frac{1}{6\sqrt{N_{q1}N_{q2}}}\sigma^{(1)}_{z}\sigma^{(2)}_{z})  \\
\end{split}
\end{equation}
where $N_{q1}$ and $N_{q2}$ are the number of levels in the potentials of qubits 1 and 2. The typical values for $N_{q1}$ and $N_{q2}$ are $N_{q1}=N_{q2}=5$.

\begin{figure}
\centering
\includegraphics[width=0.5\textwidth]{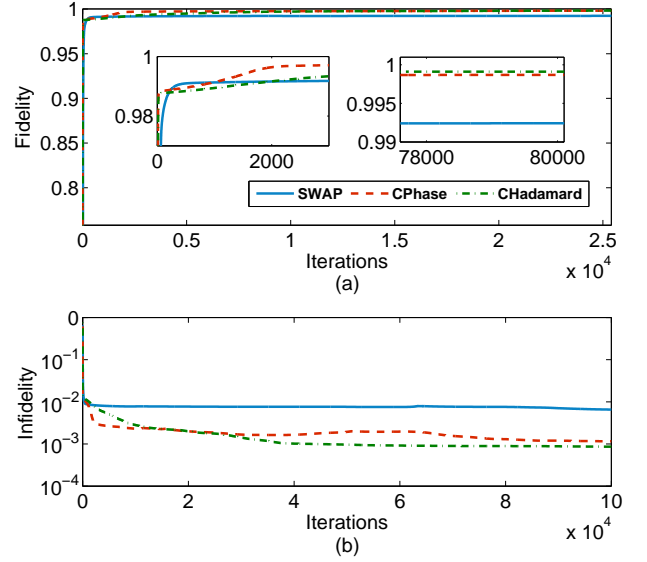}
\caption{The learning performance (a) the fidelity; (b) the infidelity versus the number of iterations for the SWAP, CPhase and CHadamard operation in the superconducting qubits.}\label{SQUID1}
\end{figure}

\begin{figure*}[!htb]
\centering
\includegraphics[width=1\textwidth]{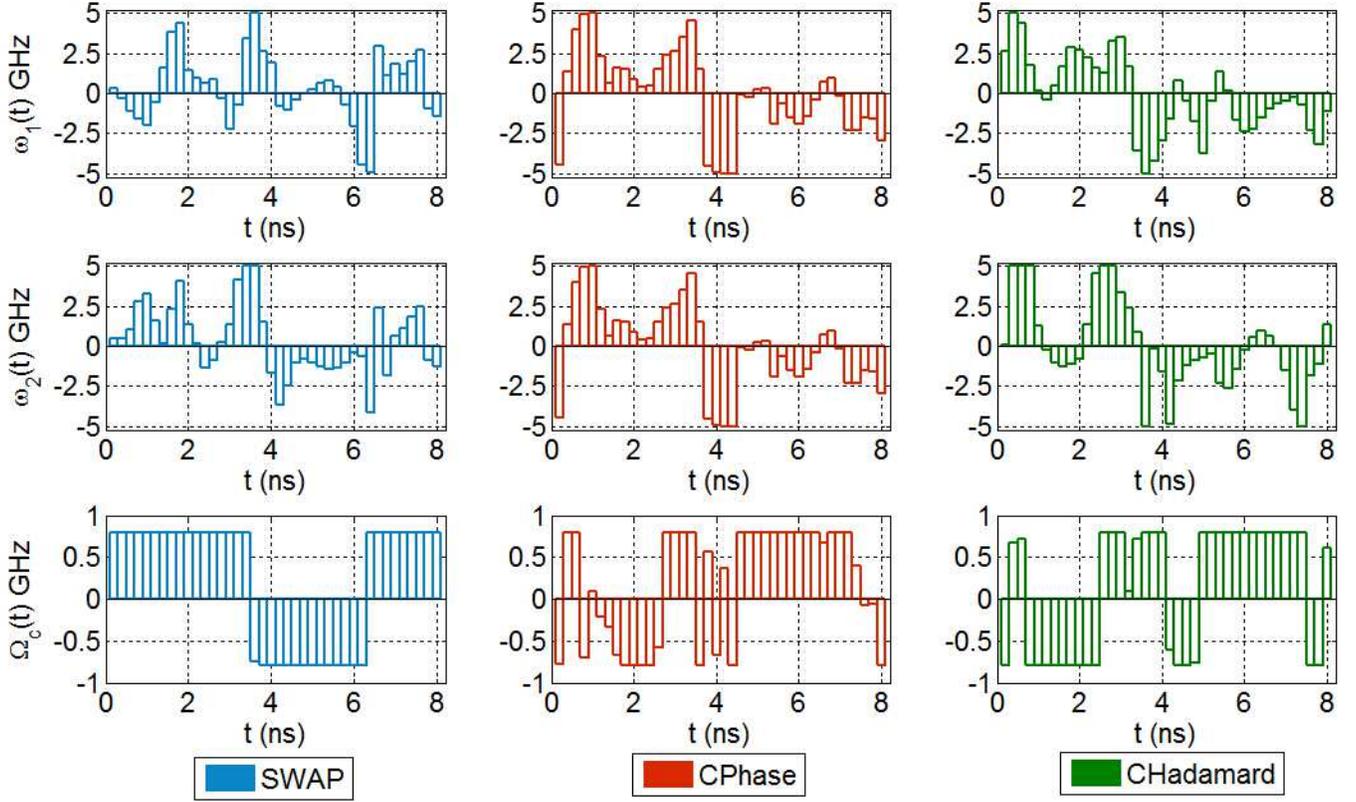}
\caption{The learned control strategy for the quantum operations in the superconducting qubits, the first, second, third column is the control fields of the SWAP, CPhase and CHadamard operation respectively.}\label{SQUID2}
\end{figure*}

\subsection{Numerical results}
We assume that the frequencies $\omega_{1}(t)$, $\omega_{2}(t) \in [-5, 5]\ \text{GHz}$ can be adjusted by changing the bias currents of two phase qubits, and $\Omega_{c}(t) \in [-800, 800]\ \text{MHz}$ can be adjusted by changing the bias current in the coupler. Let $\omega_{3}=\omega_{4}=1\ \text{GHz}$, the operation time $T=8\ \text{ns}$ and $T$ is divided into 40 smaller time intervals. The iteration step size is set as $\alpha_s=0.1$. The default initial control fields are $\omega_1=\omega_2=\text{sin}t$ \text{GHz}, $\Omega_c=0.05\text{sin}t$ \text{GHz}.

Due to possible fluctuations, we assume that the practical Hamiltonian has the following form
\begin{equation}\label{coupledHamiltonian2}
\begin{split}
H &= \frac{\hbar\varepsilon_{1}\omega_{1}(t)}{2}\sigma^{(1)}_{z}+\frac{\hbar\varepsilon_{2}\omega_{2}(t)}{2}\sigma^{(2)}_{z}
     +\frac{\hbar \omega_{3}}{2}\sigma^{(1)}_{x}+\frac{\hbar \omega_{4}}{2}\sigma^{(2)}_{x}  \\
  & \quad +\frac{\hbar\varepsilon_{3}\Omega_{c}(t)}{2}(\sigma^{(1)}_{x}\sigma^{(2)}_{x}+
    \frac{1}{30}\sigma^{(1)}_{z}\sigma^{(2)}_{z})  \\
\end{split}
\end{equation}
with $\varepsilon_{j}\in [1-E, 1+E]$ ($j=1,2,3$). We assume $E=0.1$.

Now we use the proposed method to achieve the SWAP, CPhase and CHadamard gates, respectively.
The SWAP gate is a quantum gate which swaps the states of two qubits, and it can be represented by the matrix
\begin{equation}
\text{SWAP}=\begin{pmatrix}
 \ 1 & \ 0 & \ 0 & \ 0 \ \\
 \ 0 & \ 0 & \ 1 & \ 0 \ \\
 \ 0 & \ 1 & \ 0 & \ 0 \ \\
 \ 0 & \ 0 & \ 0 & \ 1 \ \\
\end{pmatrix}.
\end{equation}
Controlled gates act on two or more qubits, where one or more qubits act as a control for some operation. For example, the controlled-Phase/Hadamard gate (or CPhase/CHadamard) acts on two qubits, and performs the Phase/Hadamard operation on the second qubit only when the first qubit is $|1\rangle$ , otherwise leaves it unchanged. They are represented by the matrices
\begin{equation}
\text{CPhase}=\begin{pmatrix}
 \ 1 & \ \ 0 & \ \ 0 & \ 0 \ \\
 \ 0 & \ \ 1 & \ \ 0 & \ 0 \ \\
 \ 0 & \ \ 0 & \ \ 1 & \ 0 \ \\
 \ 0 & \ \ 0 & \ \ 0 & \ -1 \ \\
\end{pmatrix},  \nonumber
\end{equation}
\begin{equation}
\text{CHadamard}=\begin{pmatrix}
 \ 1 \ & \ 0 & \ 0 & 0  \\
 \ 0 \ & \ 1 & \ 0 & 0  \\
 \ 0 \ & \ 0 & \ \frac{1}{\sqrt{2}} & \frac{1}{\sqrt{2}}  \\
 \ 0 \ & \ 0 & \ \frac{1}{\sqrt{2}} & -\frac{1}{\sqrt{2}}  \\
\end{pmatrix}.
\end{equation}

The result is shown in Fig. \ref{SQUID1}. In the training step, the precision of the SWAP operation can achieve slightly above $99.0\%$, and the precision of the CPhase operation and the CHadamard operation can achieve around $99.90\%$.
The learning process takes 4.2 hours using a computer with CPU@3.00GHz, Windows 7, Matlab R2013a. The learned control strategies are shown in Fig. \ref{SQUID2}, where the first, second and third columns are the control fields of the SWAP, CPhase and CHadamard operations, respectively. In the testing step, the learned fields are applied to 1000 samples which are generated randomly by selecting values of fluctuation parameters according to the uniform distribution. The average fidelity of the SWAP, CPhase and CHadamard operation reaches 0.9934, 0.9987 and 0.9991, respectively (as shown in Fig. \ref{SQUID3}), which verifies the robust realization of these operations.
These results are listed in Table \ref{table_SQUID}. It is clear that the higher fidelity of one operation achieved in the training step, the higher its average fidelity of the tested samples can be reached in the testing step.

\begin{table}[!htb] \centering
\caption{Fidelities of the three quantum operations achieved in the training/testing steps}\label{table_SQUID}
\begin{tabular}{|c|c|c|c|}
\hline Operation & SWAP & CPhase & CHadamard \\
\hline Training fidelity & 0.9935 & 0.9988 & 0.9991 \\
\hline Testing fidelity & 0.9934 & 0.9987 & 0.9991 \\
\hline
\end{tabular}
\end{table}

\begin{figure}
\centering
\includegraphics[width=0.5\textwidth]{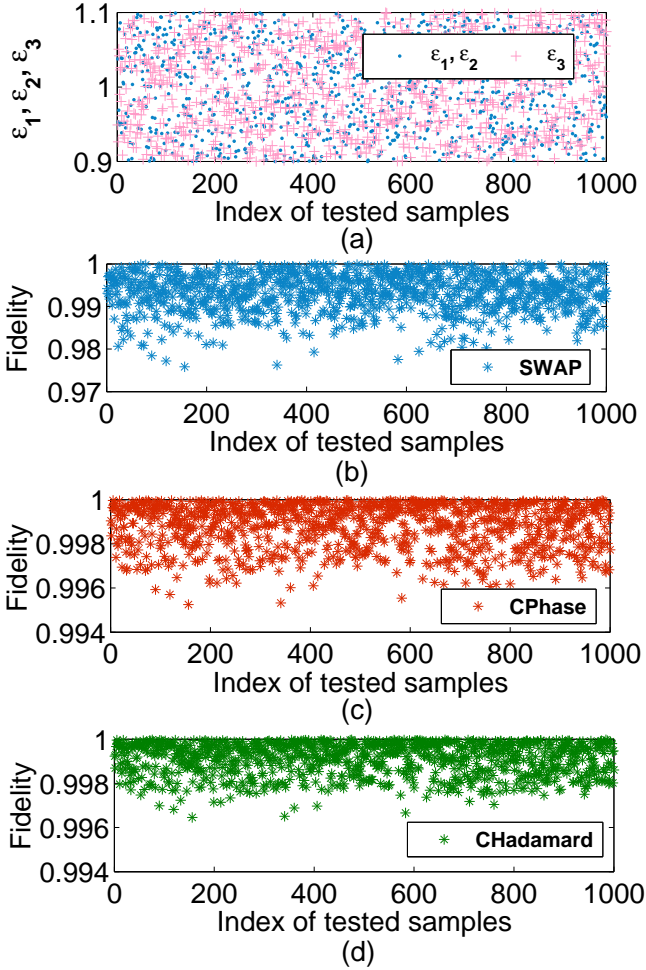}
\caption{The testing performance of the operations in the superconducting qubits: (a) 1000 randomly selected testing samples; the testing performance regarding fidelity of (b) the SWAP operation; (c) the CPhase operation; and (d) the CHadamard operation.}\label{SQUID3}
\end{figure}

\section{Robust unitary transformation in \\ quantum spin chain}\label{Sec6}
One of the admirable features of quantum technologies is its ability to establish amazing correlations between a pair of particles. In particular, spin chains with nearest-neighbor interaction have been recognized as prototypical quantum models, since they provide a wide range of solutions for quantum transformation \cite{Bose 2003PRL}, \cite{Apollaro 2013IJMPB}. In addition to their simple theoretical descriptions, they can be efficiently implemented by using arrays of trapped ions \cite{Porras 2004PRL} or optical lattices with cold atoms \cite{Winterauer 2015PRA}.
In this section, the presented SLC approach is further tested in a quantum spin chain system with uncertainties.

\subsection{The physical system}
As shown in Fig. \ref{spinchainpic}, a theoretical model of a spin chain with nearest-neighbor interactions is demonstrated with  the spin up/down in each particle representing the quantum state. The Hamiltonian of an isotropic Heisenberg spin 1/2 chain with nearest-neighbor interactions is given by
\begin{equation}
H(t)=H_0+H_c(t),
\end{equation}
where $H_0$ is the Heisenberg Hamiltonian and
\begin{equation}
H_0=J\sum_{n=1}^{N_s-1}(S_{n}^x S_{n+1}^x + S_{n}^y S_{n+1}^y + S_{n}^z S_{n+1}^z),
\end{equation}
while
\begin{equation}
H_c(t)=u_1^x(t)S_{1}^x+u_1^y(t)S_{1}^y+u_2^x(t)S_{2}^x+u_2^y(t)S_{2}^y
\end{equation}
is the control Hamiltonian. The time-dependent control fields may be applied only on the first two spins.
$N_s$ is the length of the spin chain, $S^\alpha=\sigma_\alpha/2$ ($\alpha=x, y, z$) are spin 1/2 operators, $\sigma_x, \sigma_y, \sigma_z$ are Pauli operators, and $J>0$ is the antiferromagnetic exchange interaction between spins. All frequencies and control field amplitudes can be expressed in units of the coupling strength $J$, and all times in units of $1/J$. For convenience, throughout this section we set $J=1$.
The above spin chain model is called the Heisenberg XXX model.

\begin{figure}
\centering
\includegraphics[width=0.48\textwidth]{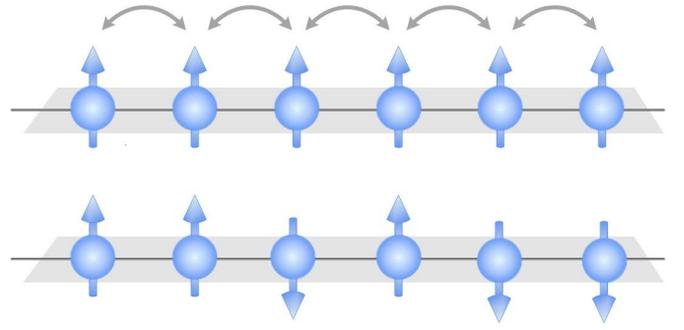}
\caption{A spin chain with nearest-neighbor interactions.}\label{spinchainpic}
\end{figure}

\begin{figure}
\centering
\includegraphics[width=0.5\textwidth]{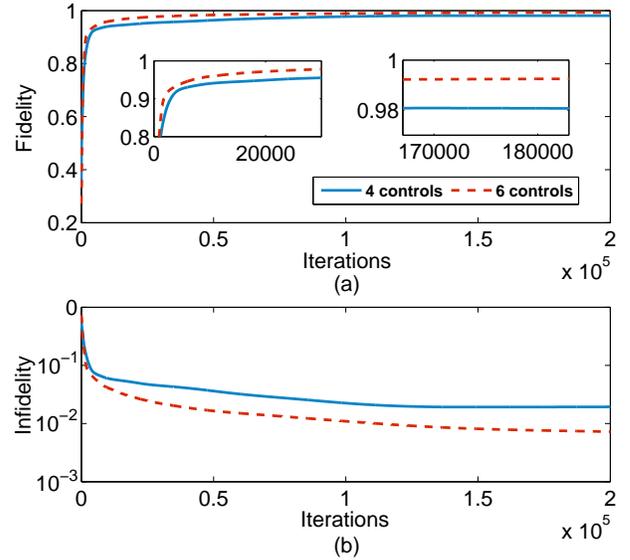}
\caption{The learning performance (a) the fidelity; (b) the infidelity versus the number of iterations for the quantum state transformation in a spin chain.}\label{spinchain1}
\end{figure}

We assume that the Hamiltonian with uncertainties can be written as
\begin{equation}\label{spinchain_uncerpara}
H(t)=\varepsilon_0 H_0+\varepsilon_c H_c(t),
\end{equation}
where $\varepsilon_0$ and $\varepsilon_c$ represent uncertainty parameters in the free Hamiltonian and control Hamiltonian, respectively. We assume that the uncertainty parameters satisfy $\varepsilon_0\in[1-E,1+E]$ and $\varepsilon_c\in[1-E,1+E]$.

One role of the Heisenberg spin chain in quantum computation is to be used to perform quantum transformations. Taking the Toffoli gate, or the Controlled-Controlled-NOT (CCNOT) gate as an example, the action of the $N_s$-qubit gate $\text{CCNOT}_{N_s}$, which performs the CCNOT operation on the last three qubits in the chain, can be defined as
\begin{equation}
\text{CCNOT}_{N_s}:= I\otimes I\otimes \ldots \otimes I\otimes \text{CCNOT}.
\end{equation}
The Toffoli gate (or CCNOT) acts on three qubits, and flips the third qubit if and only if the first and the second qubits are both at state $|1\rangle$, otherwise leaves it unchanged.
This unitary transformation can be represented by the matrix
\begin{equation}
\text{CCNOT}=\begin{pmatrix}
 \ 1 & \ 0 & \ 0 & \ 0 & \ 0 & \ 0 & \ 0 & \ 0 \  \\
 \ 0 & \ 1 & \ 0 & \ 0 & \ 0 & \ 0 & \ 0 & \ 0 \  \\
 \ 0 & \ 0 & \ 1 & \ 0 & \ 0 & \ 0 & \ 0 & \ 0 \  \\
 \ 0 & \ 0 & \ 0 & \ 1 & \ 0 & \ 0 & \ 0 & \ 0 \  \\
 \ 0 & \ 0 & \ 0 & \ 0 & \ 1 & \ 0 & \ 0 & \ 0 \  \\
 \ 0 & \ 0 & \ 0 & \ 0 & \ 0 & \ 1 & \ 0 & \ 0 \  \\
 \ 0 & \ 0 & \ 0 & \ 0 & \ 0 & \ 0 & \ 0 & \ 1 \  \\
 \ 0 & \ 0 & \ 0 & \ 0 & \ 0 & \ 0 & \ 1 & \ 0 \  \\
\end{pmatrix}.
\end{equation}
The fidelity between the quantum unitary transformation $U(T)$ and the target transformation $U_F$ can be defined as follows
\begin{equation}
F(T) = \frac{1}{2^{N_s}} |\text{tr}\{U_F^\dagger U(T)\}| ,
\end{equation}
where $N_s$ denotes the number of qubits under consideration in the spin chain.
For simplicity, we only consider the evolution of the last three qubits in the chain in this section, i.e., $N_s=3$.
During the learning step of SLC, this performance function is used to measure the fidelity of the system under a given control law. An optimal control law can be found by maximizing $F(T)$.
In this section, we set transformation time $T=20$. The iteration step size is set as $\alpha_s=0.01$.

\begin{figure}
\centering
\includegraphics[width=0.5\textwidth]{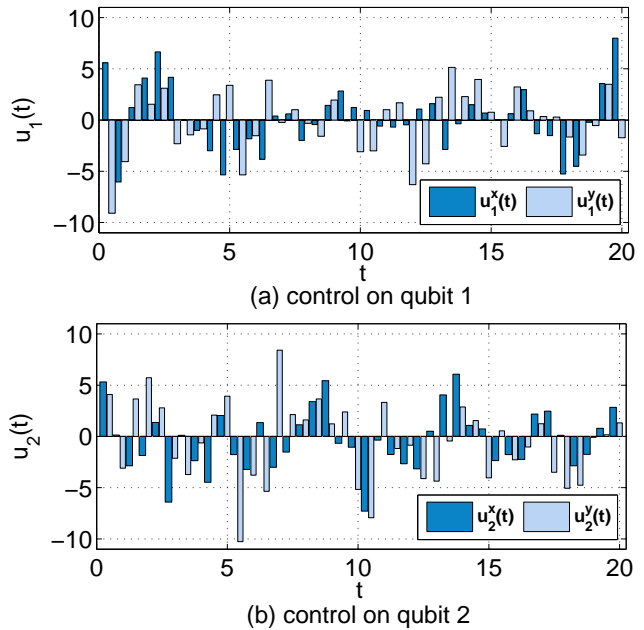}
\caption{The learned control strategy for the spin chain quantum-state-transformation with 4 control pulse sequences.}\label{spinchain2}
\end{figure}

\begin{figure}
\centering
\includegraphics[width=0.5\textwidth]{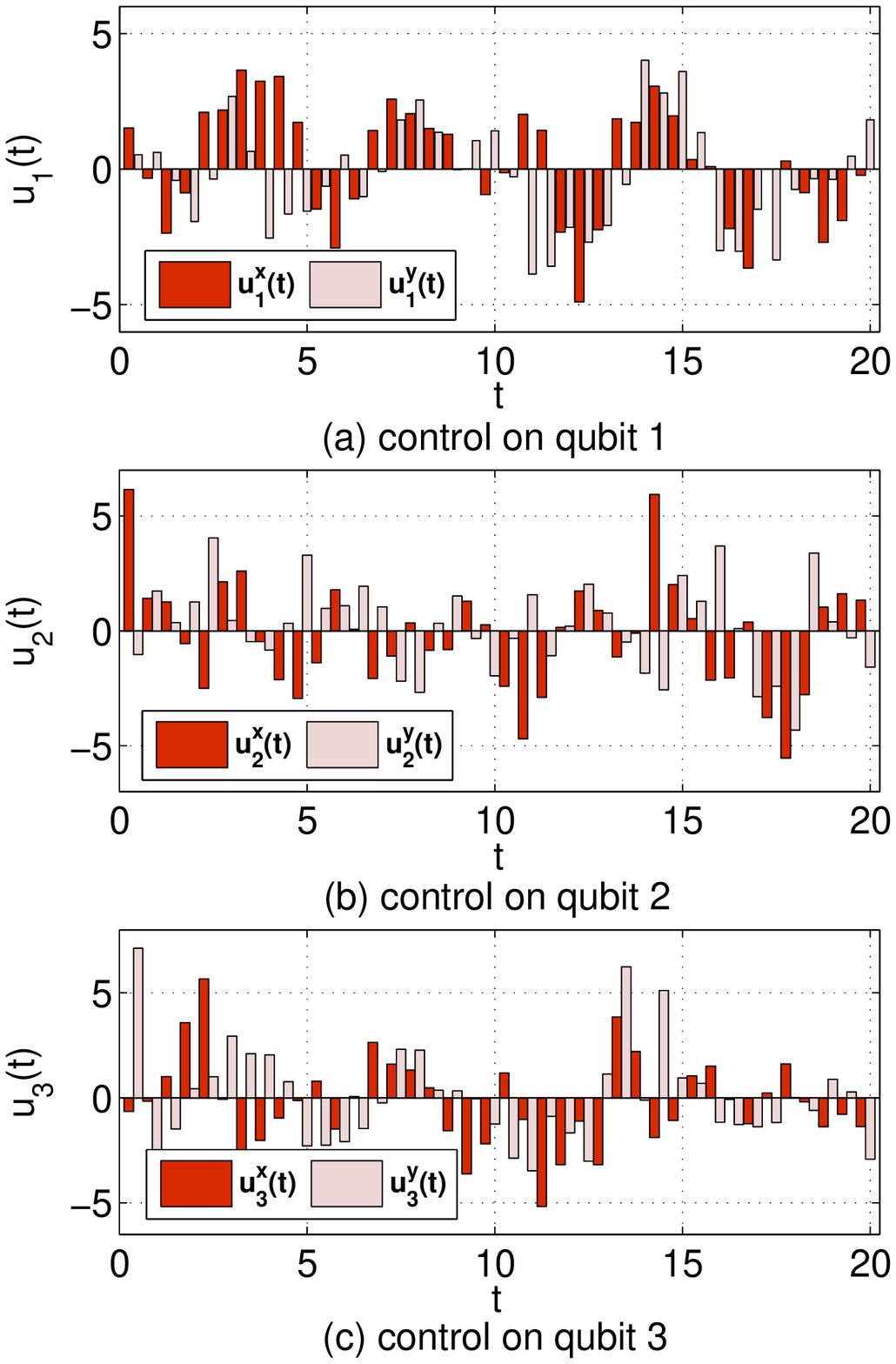}
\caption{The learned control strategy for the spin chain quantum-state-transformation with 6 control pulse sequences.}\label{spinchain3}
\end{figure}

\begin{figure}
\centering
\includegraphics[width=0.5\textwidth]{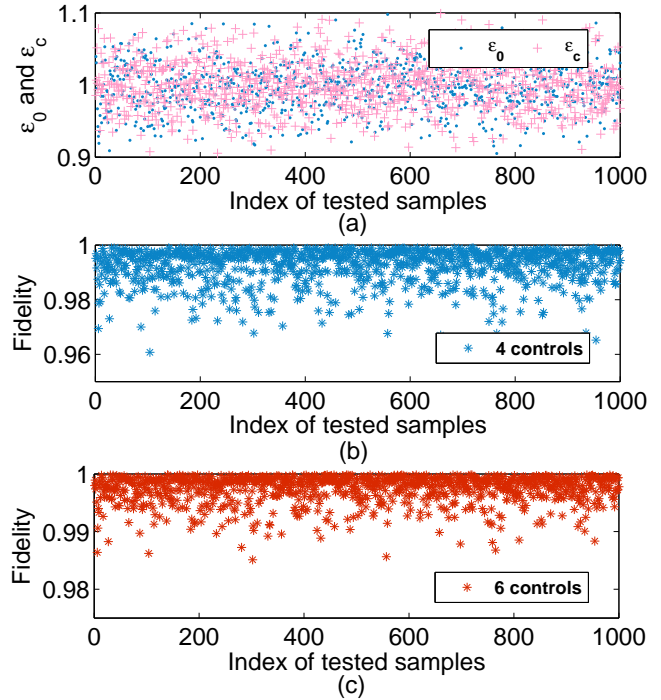}
\caption{The testing performance of the spin chain quantum-state-transformation: (a) 1000 randomly selected testing samples; the testing performance regarding fidelity of the CCNOT operation with (b) four control pulse sequences; and (c) six control pulse sequences.}\label{spinchain4}
\end{figure}

    \subsection{Numerical results}
In physical realization, in order to avoid the error arising from the interfere of two control pulses, these two control pulses may be applied to the system alternately. During the first half of the pulse duration ($\Delta t=T/N$) in the $j$th interval, we apply one $x$ control to the first spin of the chain with amplitude $u_1^x(j)$ and another $x$ control to the second spin with amplitude $u_2^x(j)$. That is, the system is governed by the Hamiltonian $H_j^x=H_0+u_1^x(j)S_1^x+u_2^x(j)S_2^x$. Subsequently we apply one $y$ control with amplitude $u_1^y(j)$ and another $y$ control with amplitude $u_2^y(j)$ to the first and the second spin of the chain, respectively, in the second half of the $j$th time interval. Therefore the system evolves under the Hamiltonian $H_j^y=H_0+u_1^y(j)S_1^y+u_2^y(j)S_2^y$.
The whole quantum unitary transformation during the operation time $T$ may be described as
\begin{equation}\label{spinchain_whole}
U(T)=U_N^y U_N^x \cdots U_j^y U_j^x \cdots U_1^y U_1^x U_0,
\end{equation}
where $U_j^x=e^{-i H_j^x \Delta t/2}$ and $U_j^y=e^{-i H_j^y \Delta t/2}$ are the half-interval unitary transformation, respectively.

In this subsection, the proposed method is applied to find a robust control for a CCNOT operation of three interacting qubits in a spin chain system.
The Hamiltonian and its parameters are presented in the previous subsection. The training performance is shown in Fig. \ref{spinchain1}. The average fidelity of the augmented system of the CCNOT operation converges to 0.9808 after 200,000 iterations. The learned control strategy is shown in Fig. \ref{spinchain2}. Then in the testing step, the learned fields are applied to 1000 randomly selected samples whose uncertainty parameters have truncated Gaussian distribution (with mean 1 and standard deviation $E/3$), and the average fidelity of the CCNOT operation reaches 0.9924 shown in Fig. \ref{spinchain4}(b), which demonstrates the robustness of our proposed method.

In practical applications, it would be more convenient if we use less number of controls. But at the same time, the fidelity of one unitary transformation we achieved may be slightly decreased with less number of controls, as well as larger control amplitudes are usually needed. In the particular quantum model presented in the previous subsection, we may use up to six control pulse sequences; that is, we may apply both $x$ and $y$ direction controls on all of the three qubits.
Under this circumstance, the control Hamiltonian may be rewritten as
\begin{equation}
H_c(t)=\sum_{i=1}^3 (u_i^x(t)S_{i}^x+u_i^y(t)S_{i}^y),
\end{equation}
and the half-interval system Hamiltonian may be rewritten as
\begin{equation}
\begin{split}
H_j^x=H_0+\sum_{i=1}^3 u_i^x(j)S_{i}^x, \\
H_j^y=H_0+\sum_{i=1}^3 u_i^y(j)S_{i}^y.
\end{split}
\end{equation}
The whole quantum unitary transformation during the operation time $T$ is the same as Eq. (\ref{spinchain_whole}).
Besides, we assume the settings of the uncertainty parameters are still the same as Eq. (\ref{spinchain_uncerpara}).

The training performance with six control pulse sequences is shown in Fig. \ref{spinchain1} (dashed line). The average fidelity of the augmented system of the CCNOT operation converges to 0.9927 after 200,000 iterations, which is better than that with four controls. The learned control strategy is shown in Fig. \ref{spinchain3}. It can be found that the control amplitudes are smaller than that with four controls.
Usually, it is easier to achieve the control objective using more control pulse sequences because we have more flexibility for control design. The maximum amplitude and the mean amplitude under these two circumstances are all listed in Table \ref{table_spinchain}.
Then in the testing step, the learned fields are applied to 1000 randomly selected samples whose uncertainty parameters have truncated Gaussian distribution (with mean 1 and standard deviation $E/3$), and the average fidelity of the CCNOT operation reaches 0.9973 (see Fig. \ref{spinchain4}(c)), which further demonstrates the robustness of the proposed method.
More detailed results are summarized in Table \ref{table_spinchain}.

\begin{table}[!htb] \centering
\caption{Fidelities of the Toffoli gate achieved in the training/testing step and the maximum/mean value of their control amplitudes under 4/6 control pulse sequences}\label{table_spinchain}
\begin{tabular}{|c|c|c|c|}
\hline \multicolumn{2}{|c|}{Number of controls} & 4 controls & 6 controls \\
\hline \multirow{2}{*}{Fidelity} & Training step & 0.9808 & 0.9927 \\
\cline{2-4} & Testing step & 0.9924 & 0.9973 \\
\hline \multirow{2}{*}{Control amplitudes} & Maximum value & 10.26 & 7.13 \\
\cline{2-4} & Mean value & 2.60 & 1.71 \\
\hline
\end{tabular}
\end{table}

All these results further prove that the higher fidelity of a unitary transformation achieved in the training step, the higher its average fidelity of the tested samples can achieve in the testing step.
By comparing Table \ref{table_SQUID} and Table \ref{table_spinchain}, it is clear that if the uncertainty parameters of the tested samples are selected according to the truncated Gaussian distribution other than the uniform distribution, better fidelities in the testing step may be achieved.

\section{Conclusion}\label{Sec7}
In this paper, an SLC approach was employed to achieve quantum optimal control laws for robust unitary transformations.
The proposed method has been applied to three typical examples of robust control problems including robust quantum transformations in a three-level quantum system, achieving robust SWAP, CPhase and CHadamard operations in a superconducting quantum circuit, and CCNOT gate in a spin chain system. Several groups of numerical results demonstrate that even when the uncertainty parameters have quite large fluctuations, the proposed SLC method is still effective for the creation of robust unitary transformations. If we could estimate the fluctuation bound better (i.e., a smaller bound), we may achieve better performance using the proposed method.
Our future work will focus on expanding the proposed method to other types of uncertainties and other tasks in quantum systems (e.g., synchronization \cite{LiuZhixin2016} and switching control \cite{Kang2016}), as well as developing more efficient and practical methods by combing other learning algorithms, such as reinforcement learning and approximate dynamic programming algorithms \cite{SMCB2008}-\cite{TNNLS2014}, genetic algorithms or differential evolution \cite{Das and Suganthan 2011}.
%%%%%%%%%%%%%%%%%%%%%%%%%%%%%%%%%%%%%%%%%%%%%%%%%%%%%%%%%%%%%%%%%%%%%%%%%%%%%%%%

\end{document}